\documentclass[
,reprint 
]{revtex4-1}

\usepackage[utf8]{inputenc}     
\usepackage[T1]{fontenc}        

\usepackage{graphicx}

\usepackage[squaren]{SIunits}
\newcommand{\SI}[2]{\unit{#1}{#2}}

\usepackage{color}              
\usepackage{xspace}             

\usepackage{ifthen}

\usepackage
{hyperref}

\begin{document}
\hyphenation{
hexa-fluoro-tetra-cyano-naphtho-quino-di-methane
}

\newcommand{\insitu}[0] {in-situ\xspace}

\newcommand{\PUNKT}[0]{\ensuremath{~.}}
\newcommand{\KOMMA}[0]{\ensuremath{~,}}

\newcommand{\todo}[1]{%
\textcolor{red}{TODO: #1}\xspace%
}

\newcommand{\K}   [1] {\SI{#1}{\kelvin}}
\newcommand{\grad}[1] {\SI{#1}{\celsius}}
\newcommand{\nm}  [1] {\SI{#1}{\nano\meter}}
\newcommand{\mm}  [1] {\SI{#1}{\milli\meter}}
\newcommand{\mr}  [1] {\SI{#1}{MR}}
\newcommand{\WcmK}[1] {\SI{#1}{\watt\per\centi\meter\,\kelvin}}

\renewcommand{\c}      [1][\empty]{\ifthenelse{\equal{#1}{\empty}} {\ensuremath{\sigma}\xspace}     {\mbox{\ensuremath{\sigma=\Scm{#1}}}}}

\newcommand  {\thcond} [1][\empty]{\ifthenelse{\equal{#1}{\empty}} {\ensuremath{\kappa}\xspace}     {\mbox{\ensuremath{\kappa=\WcmK{#1}}}}}

\renewcommand{\S}      [1][\empty]{\ifthenelse{\equal{#1}{\empty}} {\ensuremath{S}\xspace}          {\mbox{\ensuremath{S=\uVK{#1}}}}}
\newcommand{\C}      [1][\empty]{\ifthenelse{\equal{#1}{\empty}} {\ensuremath{C}\xspace}          {\mbox{\ensuremath{C=\mr{#1}}}}}

\newcommand  {\T}     [1][\empty]{\ifthenelse{\equal{#1}{\empty}} {\ensuremath{T}\xspace} {\mbox{\ensuremath{T=\grad{#1}}}}}
\newcommand  {\Tm}     [1][\empty]{\ifthenelse{\equal{#1}{\empty}} {\ensuremath{T_\text{M}}\xspace} {\mbox{\ensuremath{T_\text{M}=\grad{#1}}}}}
\newcommand  {\Td}     [1][\empty]{\ifthenelse{\equal{#1}{\empty}} {\ensuremath{\Delta T}\xspace}   {\mbox{\ensuremath{\Delta T=\K{#1}}}}}
\newcommand  {\Th}     [1][\empty]{\ifthenelse{\equal{#1}{\empty}} {\ensuremath{T_\text{H}}\xspace} {\mbox{\ensuremath{T_\text{H}=\grad{#1}}}}}
\newcommand  {\Tc}     [1][\empty]{\ifthenelse{\equal{#1}{\empty}} {\ensuremath{T_\text{C}}\xspace} {\mbox{\ensuremath{T_\text{C}=\grad{#1}}}}}
\newcommand  {\ZTm}    [1][\empty]{\ifthenelse{\equal{#1}{\empty}} {\ensuremath{Z\Tm}\xspace}       {\mbox{\ensuremath{Z\Tm=#1}}}}

\renewcommand{\C}      [1][\empty]{\ifthenelse{\equal{#1}{\empty}} {\ensuremath{C}\xspace}          {\mbox{\ensuremath{C=\mr{#1}}}}}
\newcommand  {\V}      [1][\empty]{\ifthenelse{\equal{#1}{\empty}} {\ensuremath{V}\xspace}          {\mbox{\ensuremath{V=\SI{#1}{\volt}}}}}
\newcommand  {\eff}    [1][\empty]{\ifthenelse{\equal{#1}{\empty}} {\ensuremath{\eta}\xspace}       {\mbox{\ensuremath{\eta=#1\%}}}}

\newcommand{\Eqnref}[1] {Equation~(\ref{eq:#1})}
\newcommand{\eqnref}[1] {equation~(\ref{eq:#1})}
\newcommand{\eqref}[1] {(\ref{eq:#1})}
\newcommand{\Figref}[1] {FIG.~\ref{fig:#1}}
\newcommand{\figref}[1] {FIG.~\ref{fig:#1}}

\newcommand{\meo} [0] {\mbox{MeO-TPD}\xspace}
\newcommand{\meoLong} [0] {\textit{N},\textit{N},\textit{N}',\textit{N}'-tetrakis 4-methoxyphenyl-benzidine\xspace}

\newcommand{\lili} [0] {\mbox{BF-DPB}\xspace}
\newcommand{\liliLong} [0] {\textit{N},\textit{N}'-bis(9,9-dimethyl-fluoren-2-yl)-\textit{N},\textit{N}'-diphenyl-benzidine\xspace}
\newcommand{\CSF} [0] {\texorpdfstring{C$_{60}$F$_{36}$}{C60F36}\xspace}
\newcommand{\CSFLong} [0] {36-fold fluorinated \texorpdfstring{C$_{60}$}{C60}\xspace}
\newcommand{\FS} [0] {\texorpdfstring{\mbox{F$_{6}$-TCNNQ}}{F6-TCNNQ}\xspace}
\newcommand{\FSLong}[0] {1,3,4,5,7,8-hexafluorotetracyanonaphthoquinodimethane\xspace}
\newcommand{\FV} [0] {\texorpdfstring{\mbox{F$_{4}$-TCNQ}}{F4-TCNQ}\xspace}
\newcommand{\FVLong}[0] {tetrafluoro-tetracyanoquinodimethane\xspace}
\newcommand{\CS} [0] {\texorpdfstring{C$_{60}$}{C60}\xspace}
\newcommand{\CrPd}[0] {\texorpdfstring{Cr$_2$(hpp)$_4$}{Cr2(hpp)4)}\xspace}
\newcommand{\WPd}[0] {\texorpdfstring{W$_2$(hpp)$_4$}{W2(hpp)4)}\xspace}
\newcommand{\CrPdLong}[0] {tetrakis(1,3,4,6,7,8-hexahydro-2H-pyrimido[1,2-a]pyrimidinato)\-dichromium\,(II)\xspace}
\newcommand{\WPdLong}[0]  {tetrakis(1,3,4,6,7,8-hexahydro-2H-pyrimido[1,2-a]pyrimidinato)\-ditungsten\,(II)\xspace}
\newcommand{\aob} [0] {AOB\xspace}
\newcommand{\aobLong} [0] {3,6-bis(dimethylamino)acridine\xspace}
\newcommand{\dmbiPOH}[0] {\mbox{DMBI-POH}\xspace}
\newcommand{\dmbi}[0] {\dmbiPOH}
\newcommand{\dmbiPOHLong}[0] {2-(1,3-dimethyl-1\textsl{H}-benzoimidazol-3-ium-2-yl)phenolatehydrate\xspace}
\newcommand{\meodmbiI}[0] {\mbox{\textsl{o}-MeO-DMBI-I}\xspace}
\newcommand{\meodmbiILong}[0] {2-(2-methoxyphenyl)-1,3-dimethyl-1\textsl{H}-benzoimidazol-3-ium iodide\xspace}
\renewcommand{\todo}[1]{}

\title{Thermoelectric properties of doped small molecule organic semiconductor films}
\begin{abstract}
\noindent
Doped films of organic small molecules are investigated with respect to their thermoelectric properties. A variety of hosts and dopants, for both $n$ and $p$-doping, are compared. \CS $n$-doped by \CrPd or \meodmbiI are found to be the most promising material systems with a maximum of \ZTm[0.069] at \Tm[40], assuming a doping-independent thermal conductivity due to phonon-based heat transport.
This value is 16\% of the current record reported for optimized devices employing the doped polymer \mbox{PEDOT:PSS}.
\end{abstract}

\newcommand{\affIapp}   [0] {Institut f\"ur Angewandte Photophysik, Technische Universit\"at Dresden, 01062 Dresden, Germany, http://www.iapp.de}
\newcommand{\affOxford} [0] {Physics Department, University of Oxford, Clarendon Laboratory, Parks Road, Oxford OX1 3PU, England, United Kingdom}

\author{Torben Menke}
\email{torben.menke@iapp.de}
\affiliation{\affIapp}

\maketitle

\section{Introduction}
The thermoelectric (TE) effect allows for generation of electricity from temperature gradients. This opens the path for self-sustaining sensors, power generation from waste heat or a complementary to photovoltaics\cite{Kraemer2011}.
In recent years, several approaches for different material systems and device layouts have been developed to achieve a high efficiency at low cost.\cite{Tritt2011} While the highest efficiencies have been reported for TE elements based on inorganic semiconductors, recently, doped polymers gained attention because despite having lower efficiency they have the potential of a strong cost reduction.\cite{Kim2013,Bubnova2013} Additionally, these organic compounds allow for fabrication of TE devices on flexible foils which are easier to handle and can e.g.\ be installed on large and curved surfaces like in cooling towers.

In this work, another class of organic semiconducting compounds is studied: small molecules, which so far have hardly been examined with respect to their TE properties, but are predicted to be a promising material class\cite{Chen2012,Wang2012}.
Small molecules have the advantage to polymers that due to their low molecular weight they can be deposited by thermal evaporation in vacuum, allowing for high material purity and arbitrary layer stacks, while mass production on large substrates via roll-to-roll fabrication is feasible as well. Doping of small molecules has led to novel devices like organic light-emitting diodes (OLEDs) and organic photovoltaic cells.\cite{LuessemRiedeLeo2013-PSS}
The electrical conductivity of such doped layers of small molecules typically increases with temperature\cite{Fujimori1994,Pfeiffer1998}, while their thermal conductivity is rather low\cite{Olson1993}\todo{better source}. The scope of this publication is the comparison of the TE potential of several material combinations of small molecular hosts and dopants.

It is found that $n$-doped \CS samples show several orders of magnitude higher power factors, compared to $p$-doped hole transporting compounds. The highest values are obtained for low concentrations of \CrPd or high concentrations of \meodmbiI in \CS. The resulting \ZTm of these first tests for this material class are as high as 16\% of the current record reported for optimized devices employing the doped polymer \mbox{PEDOT:PSS}.

\todo{add further summary of this paper's results?}

\begin{figure*}[t]%
\centering%
\includegraphics{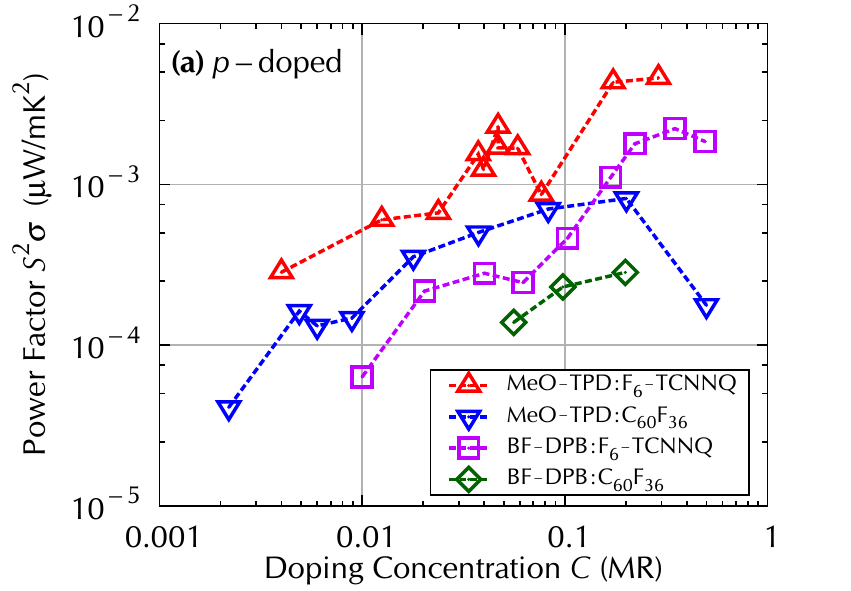}%
\includegraphics{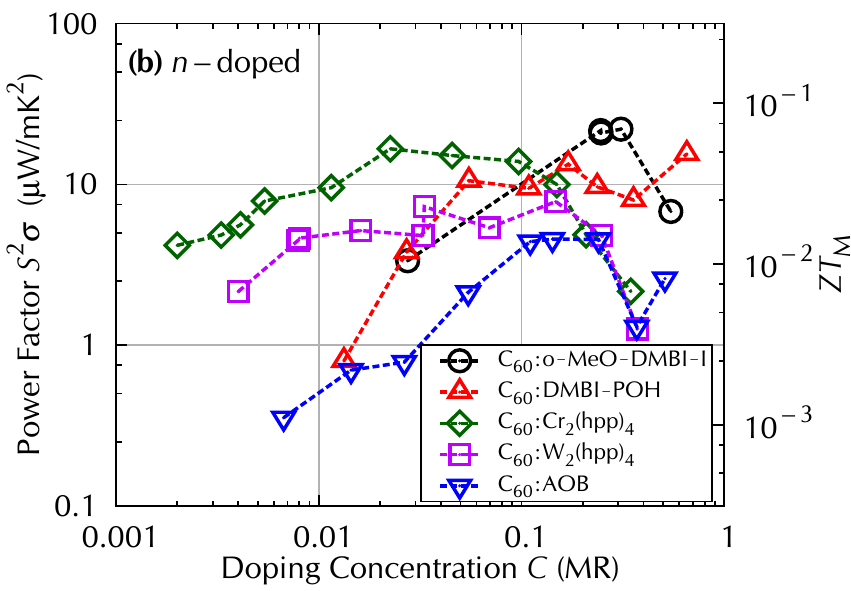}%
\caption{Power factor $\S^2\c$, at \Tm[40], \Td[5] for (a) $p$-doped and (b) $n$-doped samples. For the $n$-\CS samples the intrinsic thermal conductivity \thcond[10^{-3}]\cite{Olson1993} is used to calculate \ZTm which is displayed at the right hand side axes.
}%
\label{fig:fig1-ZT-vs-C}%
\end{figure*}%

\section{Experimental}
Samples are prepared and measured \insitu in a vacuum setup (base pressure $\approx 3\times10^{-5}$\,Pa), which has been described in detail earlier\cite{Menke2012a,TorbenMenkeDiss}. A doped organic layer of mostly \nm{30} thickness is deposited by thermal co-deposition of host and dopant onto a glass substrate with two parallel, \nm{40} thick gold contacts of \mm{5} distance. This sample layout allows for measurement of both, conductivity and Seebeck coefficient at different temperatures, because the substrate is mounted onto two separately heated copper blocks below the gold contacts. 
All samples are thermally annealed at \T[100] in vacuum prior to investigation to ensure reproducibility of temperature-dependent measurements.

Electrical measurements are performed using a Keithley 236 source measure unit. For probing the conductivity \c, a voltage of \V[1] is applied to the gold contacts and the current flow is measured and averaged over 2 minutes to compensate for statistical fluctuations. As for all samples a linear and symmetric current response is found for $\V=-10$ to \SI{10}{\volt}, we assume ohmic injection. For measuring the Seebeck coefficient \S, a temperature difference of \Td[5] between the gold contacts is used and the thermovoltage is measured and averaged for several minutes, followed by a measurement at \Td[-5] to compensate for measurement artifacts. Both, \c and \S, are measured at different mean temperature \Tm and for various samples of different doping concentration \C. In the following, \C is expressed in terms of molar ratio (MR), being the ratio of the numbers of dopant to host molecules.

Hole transporting materials investigated here are the hosts \meoLong (\meo) and \liliLong (\lili) $p$-doped by \FSLong (\FS) and \CSFLong (\CSF).\cite{Menke2013} 
As electron transporting material the host \CS is studied, comparing 5 $n$-dopants:
\CrPdLong and -ditungsten (II) (\CrPd\cite{Menke2012} and \WPd\cite{Menke2012})
, \aobLong (\aob\cite{Menke2012a})
, \dmbiPOHLong (\dmbiPOH\cite{Menke2012a}) and 
\meodmbiILong (\meodmbiI\cite{Wei2012,TorbenMenkeDiss}).
Some of the conductivity and Seebeck data presented in this work have been published earlier in the given references, but so far no conclusion for the TE properties have been deduced.

\section{Results}
The thermoelectric potential of a material is determined by the material's figure of merit \ZTm
\begin{equation}
\label{eq:ZTm}
\ZTm=\frac{\S^2 \cdot \c}{\thcond} \cdot \Tm
\KOMMA
\end{equation}
with Seebeck coefficient \S, electrical conductivity \c and thermal conductivity \thcond, measured at the mean temperature \Tm. 
Hence, for TE applications, materials of low \thcond but high \S and \c are desired.
The numerator $\S^2\c$ in \eqnref{ZTm} is called power factor.  
Doping a semiconducting TE material allows for manipulating these parameters. The presence of dopants leads to an increasing density of free charge carriers and thus \c along with a reduction of \S, due to the shift of the Fermi level towards the transport level\cite{Fritzsche1971}. 
At low doping concentrations, the thermal conductivity \thcond is attributed to phonons while at high \C contributions from the electrons dominate.
Consequently, the optimum in doping concentration needs to be determined.

For the most extensively studied inorganic TE material system Bi$_2$Te$_3$, record values in the order of \ZTm[1.0] at \Tm[400] were achieved\cite{Tritt2011}, while for the doped polymer \mbox{PEDOT:PSS} recently a promising record of \ZTm[0.42] at room temperature has been reported\cite{Kim2013}.

The figure of merit largely affects the efficiency of power generation \eff, which is given by the ratio of electrical energy $W$ generated to the net heat flow rate $Q_\text{H}$\cite{Sherman1960,Snyder2003}: 
\begin{equation}\label{eq:eff}
\eff=\frac{W}{Q_\text{H}} = \frac{\Th-\Tc}{\Th} \frac{\sqrt{1+\ZTm}-1}{\sqrt{1+\ZTm} + \Tc/\Th}
\KOMMA
\end{equation}
were \Th and \Tc are the temperatures at the hot and cold side of the material, respectively. 
For $\ZTm\to\infty$, \eqnref{eff} approaches the Carnot limit.

Layers of doped small organic molecules have been found to show an increasing \c with temperature in the range of 30 to \grad{100}, while \S stays almost constant \cite{TorbenMenkeDiss}. These properties make this class of materials a promising candidates for TE application.

Unfortunately, \thcond of such doped layers is in most cases unknown and difficult to measure, due to degradation effects by air-exposure\cite{Fujimori1994,TietzeMenke2013}. Hence, instead of \ZTm, the power factor $\S^2\c$ is used.
This evaluation is performed for a variety of material combinations and doping concentrations. Data measured at \Tm[40] are chosen, because that temperature was most reliable to control for all samples.

All four $p$-doped material combinations displayed in \figref{fig1-ZT-vs-C}~(a) show an increasing power factor with doping concentration and a saturation at high \C. The largest values are obtained for \meo doped by \FS with a maximum of \SI{4.6\cdot10^{-3}}{\micro\watt\per\meter\,\kelvin\squared}.
Several orders of magnitude higher values are obtained for $n$-doped \CS samples, as plotted in \figref{fig1-ZT-vs-C}~(b).
This is attributed to the extraordinary high electron mobility of \CS\cite{Itaka2006}, determining the conductivity. 
Hence, \CS is the most promising candidate for an organic small molecular TE device of all studied host materials. 
Applying the dopant \meodmbiI results in the highest power factors with a maximum of \SI{22}{\micro\watt\per\meter\,\kelvin\squared} at \C[0.310] ($=9.8\text{wt\%}$). 
At lower concentrations the dopant \CrPd results in the best power factors.
Interestingly, the dopants \CrPd, \WPd and \aob yield almost identical values at $\C>\mr{0.200}$.

The highest value obtained for \CS $n$-doped by \meodmbiI is comparable to the record reported for layers of \CS $n$-doped by the inorganic dopant Cs$_2$CO$_3$ in an optimized sample geometry\cite{SuminoHarada2011}. 
For the purely organic doped layers of Pentacene $p$-doped by \FV, the current record reported in literature\cite{Harada2010} is one order of magnitude lower, despite employing an optimized geometry.

\begin{figure}[t]%
\centering%
\includegraphics{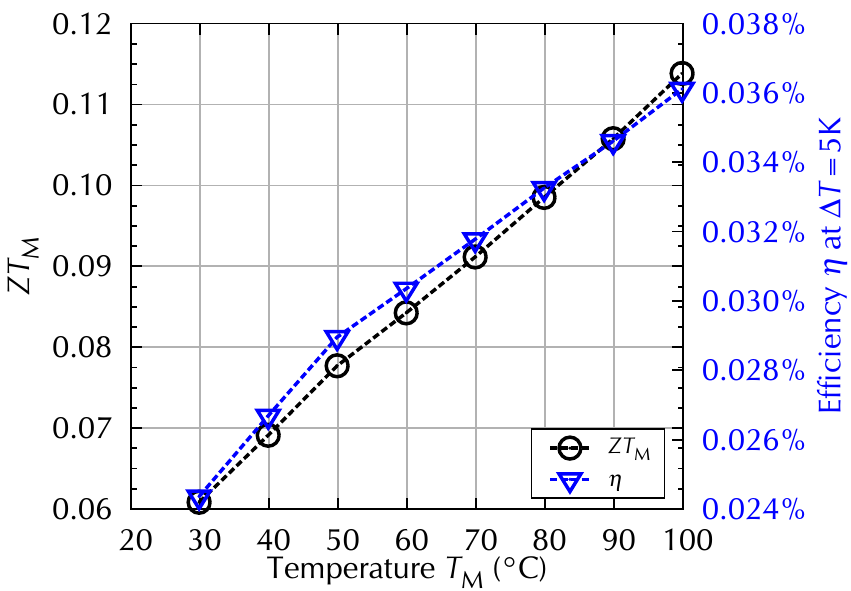}
\caption{\ZTm and \eff vs.\ \Tm at \Td[5] for a sample of \CS $n$-doped by \mr{0.310} of \meodmbiI.}%
\label{fig:eta-vs-T-106}%
\end{figure}%

For intrinsic \CS, the thermal conductivity has been reported to be almost temperature-independent between 30 and \K{300} at a value around \thcond[10^{-3}]\cite{Olson1993}, which is 4 orders of magnitude lower than values observed for \mbox{PEDOT:PSS}\cite{FengXing2008}.
Assuming that \thcond is mainly attributed to phonons and hence unaltered by introduction of dopant molecules, the value for intrinsic \CS can be used to derive \ZTm, which is shown by the right hand axes of \figref{fig1-ZT-vs-C}\,(b). A maximum of \ZTm[0.069] at \Tm[40] is found for the sample of \CS $n$-doped by \mr{0.310} of \meodmbiI. This value is 16\% of the current record reported for \mbox{PEDOT:PSS} of \ZTm[0.42] at room temperature\cite{Kim2013}.

As \c was found to increase with temperature for all investigated samples, it is interesting to study the temperature dependencies of \ZTm (\eqnref{ZTm}) and \eff (\eqnref{eff}) for the sample of highest \ZTm. Here, the temperatures \Th and \Tc, are 
\mbox{$\Tm\pm\Td/2$} \mbox{$=\Tm\pm2.5\,\text{K}$}, respectively, used during the Seebeck measurements \todo{is this correct to do?}.
\Figref{eta-vs-T-106} shows that both \ZTm and \eff show an approximately linear increase with \Tm.
Raising \Tm from \grad{30} to \grad{100}, \ZTm almost doubles while \eff rises by 50\%. At a typical TE device operating temperature of \Tm[60], the values of \ZTm[0.084] and \eff[0.030] are derived.

The efficiency reported here is rather low, compared to optimized polymer devices. 
Based on \eqnref{eff}, it is expected that increasing \Td should result in an efficiency gain. For example using $\Th=\grad{100}$ and $\Tc=\grad{20}$, hence \Td[80] instead of \Td[5] while keeping \Tm[60] and hence \ZTm[0.084], should result in a 16-fold efficiency gain, reaching \eff[0.606].
However, since in our setup it was impossible to measure at such a large \Td, this work requires a different experimental approach and will be done in future.

\section{Conclusions}

This simple estimation of the thermoelectric properties of doped organic small molecules yields that the studied $p$-doped material systems generates rather low power factors, while $n$-doped \CS is a promising candidate for organic TE devices. The best results are obtained for \CS doped by low \CrPd or high \meodmbiI doping concentrations. Assuming the thermal conductivity of intrinsic \CS is valid for doped layers as well, a maximum of \ZTm[0.069] at \Tm[40] is derived, being 16\% of the current record reported for optimized \mbox{PEDOT:PSS} layers.
\ZTm and hence \eff is found to increase with \Tm for this material class. It is expected that applying a larger temperature difference of \Td[80] at \Tm[60] an efficiency of 0.606\% can be reached. In a future work, the thermal conductivity of $n$-doped \CS layers needs to be studied to verify these results.

\todo{WT: Und nu? Kann man ZT weiter erhöhen oder wars das? Wenn ja, wie? In der Diskussion müsste mehr die thermische Leitfähigkeit von C60 diskutiert werden. Exp. dazu wären natürlich erforderlich. Wenn das Potential für n-C60 noch nicht ausgeschöpft ist, könnte ich mir vorstellen, dass das Paper wirklich Leute interessieren wird.}

\section*{Acknowledgments}
The author thanks Karl Leo and Hans Kleemann for fruitful discussions and Novaled AG, Germany for supplying the dopants \FS, \CrPd and \WPd.
\todo{plural if more than one author}


\begin{thebibliography}{23}%
\makeatletter
\providecommand \@ifxundefined [1]{%
 \@ifx{#1\undefined}
}%
\providecommand \@ifnum [1]{%
 \ifnum #1\expandafter \@firstoftwo
 \else \expandafter \@secondoftwo
 \fi
}%
\providecommand \@ifx [1]{%
 \ifx #1\expandafter \@firstoftwo
 \else \expandafter \@secondoftwo
 \fi
}%
\providecommand \natexlab [1]{#1}%
\providecommand \enquote  [1]{``#1''}%
\providecommand \bibnamefont  [1]{#1}%
\providecommand \bibfnamefont [1]{#1}%
\providecommand \citenamefont [1]{#1}%
\providecommand \href@noop [0]{\@secondoftwo}%
\providecommand \href [0]{\begingroup \@sanitize@url \@href}%
\providecommand \@href[1]{\@@startlink{#1}\@@href}%
\providecommand \@@href[1]{\endgroup#1\@@endlink}%
\providecommand \@sanitize@url [0]{\catcode `\\12\catcode `\$12\catcode
  `\&12\catcode `\#12\catcode `\^12\catcode `\_12\catcode `\%12\relax}%
\providecommand \@@startlink[1]{}%
\providecommand \@@endlink[0]{}%
\providecommand \url  [0]{\begingroup\@sanitize@url \@url }%
\providecommand \@url [1]{\endgroup\@href {#1}{\urlprefix }}%
\providecommand \urlprefix  [0]{URL }%
\providecommand \Eprint [0]{\href }%
\providecommand \doibase [0]{http://dx.doi.org/}%
\providecommand \selectlanguage [0]{\@gobble}%
\providecommand \bibinfo  [0]{\@secondoftwo}%
\providecommand \bibfield  [0]{\@secondoftwo}%
\providecommand \translation [1]{[#1]}%
\providecommand \BibitemOpen [0]{}%
\providecommand \bibitemStop [0]{}%
\providecommand \bibitemNoStop [0]{.\EOS\space}%
\providecommand \EOS [0]{\spacefactor3000\relax}%
\providecommand \BibitemShut  [1]{\csname bibitem#1\endcsname}%
\let\auto@bib@innerbib\@empty
\bibitem [{\citenamefont {Kraemer}\ \emph {et~al.}(2011)\citenamefont
  {Kraemer}, \citenamefont {Poudel}, \citenamefont {Feng}, \citenamefont
  {Caylor}, \citenamefont {Yu}, \citenamefont {Yan}, \citenamefont {Ma},
  \citenamefont {Wang}, \citenamefont {Wang}, \citenamefont {Muto},
  \citenamefont {McEnaney}, \citenamefont {Chiesa}, \citenamefont {Ren},\ and\
  \citenamefont {Chen}}]{Kraemer2011}%
  \BibitemOpen
  \bibfield  {author} {\bibinfo {author} {\bibfnamefont {D.}~\bibnamefont
  {Kraemer}}, \bibinfo {author} {\bibfnamefont {B.}~\bibnamefont {Poudel}},
  \bibinfo {author} {\bibfnamefont {H.-P.}\ \bibnamefont {Feng}}, \bibinfo
  {author} {\bibfnamefont {J.~C.}\ \bibnamefont {Caylor}}, \bibinfo {author}
  {\bibfnamefont {B.}~\bibnamefont {Yu}}, \bibinfo {author} {\bibfnamefont
  {X.}~\bibnamefont {Yan}}, \bibinfo {author} {\bibfnamefont {Y.}~\bibnamefont
  {Ma}}, \bibinfo {author} {\bibfnamefont {X.}~\bibnamefont {Wang}}, \bibinfo
  {author} {\bibfnamefont {D.}~\bibnamefont {Wang}}, \bibinfo {author}
  {\bibfnamefont {A.}~\bibnamefont {Muto}}, \bibinfo {author} {\bibfnamefont
  {K.}~\bibnamefont {McEnaney}}, \bibinfo {author} {\bibfnamefont
  {M.}~\bibnamefont {Chiesa}}, \bibinfo {author} {\bibfnamefont
  {Z.}~\bibnamefont {Ren}}, \ and\ \bibinfo {author} {\bibfnamefont
  {G.}~\bibnamefont {Chen}},\ }\href {\doibase 10.1038/nmat3013} {\bibfield
  {journal} {\bibinfo  {journal} {Nature Materials}\ }\textbf {\bibinfo
  {volume} {10}},\ \bibinfo {pages} {532} (\bibinfo {year} {2011})}\BibitemShut
  {NoStop}%
\bibitem [{\citenamefont {Tritt}\ and\ \citenamefont
  {Subramanian}(2011)}]{Tritt2011}%
  \BibitemOpen
  \bibfield  {author} {\bibinfo {author} {\bibfnamefont {T.~M.}\ \bibnamefont
  {Tritt}}\ and\ \bibinfo {author} {\bibfnamefont {M.~A.}\ \bibnamefont
  {Subramanian}},\ }\href {\doibase 10.1557/mrs2006.44} {\bibfield  {journal}
  {\bibinfo  {journal} {MRS Bulletin}\ }\textbf {\bibinfo {volume} {31}},\
  \bibinfo {pages} {188} (\bibinfo {year} {2011})}\BibitemShut {NoStop}%
\bibitem [{\citenamefont {Kim}\ \emph {et~al.}(2013)\citenamefont {Kim},
  \citenamefont {Shao}, \citenamefont {Zhang},\ and\ \citenamefont
  {Pipe}}]{Kim2013}%
  \BibitemOpen
  \bibfield  {author} {\bibinfo {author} {\bibfnamefont {G.-H.}\ \bibnamefont
  {Kim}}, \bibinfo {author} {\bibfnamefont {L.}~\bibnamefont {Shao}}, \bibinfo
  {author} {\bibfnamefont {K.}~\bibnamefont {Zhang}}, \ and\ \bibinfo {author}
  {\bibfnamefont {K.~P.}\ \bibnamefont {Pipe}},\ }\href {\doibase
  10.1038/nmat3635} {\bibfield  {journal} {\bibinfo  {journal} {Nature
  Materials}\ }\textbf {\bibinfo {volume} {12}},\ \bibinfo {pages} {719}
  (\bibinfo {year} {2013})}\BibitemShut {NoStop}%
\bibitem [{\citenamefont {Bubnova}\ \emph {et~al.}(2013)\citenamefont
  {Bubnova}, \citenamefont {Khan}, \citenamefont {Wang}, \citenamefont {Braun},
  \citenamefont {Evans}, \citenamefont {Fabretto}, \citenamefont
  {Hojati-Talemi}, \citenamefont {Dagnelund}, \citenamefont {Arlin},
  \citenamefont {Geerts}, \citenamefont {Desbief}, \citenamefont {Breiby},
  \citenamefont {Andreasen}, \citenamefont {Lazzaroni}, \citenamefont {Chen},
  \citenamefont {Zozoulenko}, \citenamefont {Fahlman}, \citenamefont {Murphy},
  \citenamefont {Berggren},\ and\ \citenamefont {Crispin}}]{Bubnova2013}%
  \BibitemOpen
  \bibfield  {author} {\bibinfo {author} {\bibfnamefont {O.}~\bibnamefont
  {Bubnova}}, \bibinfo {author} {\bibfnamefont {Z.~U.}\ \bibnamefont {Khan}},
  \bibinfo {author} {\bibfnamefont {H.}~\bibnamefont {Wang}}, \bibinfo {author}
  {\bibfnamefont {S.}~\bibnamefont {Braun}}, \bibinfo {author} {\bibfnamefont
  {D.~R.}\ \bibnamefont {Evans}}, \bibinfo {author} {\bibfnamefont
  {M.}~\bibnamefont {Fabretto}}, \bibinfo {author} {\bibfnamefont
  {P.}~\bibnamefont {Hojati-Talemi}}, \bibinfo {author} {\bibfnamefont
  {D.}~\bibnamefont {Dagnelund}}, \bibinfo {author} {\bibfnamefont {J.-B.}\
  \bibnamefont {Arlin}}, \bibinfo {author} {\bibfnamefont {Y.~H.}\ \bibnamefont
  {Geerts}}, \bibinfo {author} {\bibfnamefont {S.}~\bibnamefont {Desbief}},
  \bibinfo {author} {\bibfnamefont {D.~W.}\ \bibnamefont {Breiby}}, \bibinfo
  {author} {\bibfnamefont {J.~W.}\ \bibnamefont {Andreasen}}, \bibinfo {author}
  {\bibfnamefont {R.}~\bibnamefont {Lazzaroni}}, \bibinfo {author}
  {\bibfnamefont {W.~M.}\ \bibnamefont {Chen}}, \bibinfo {author}
  {\bibfnamefont {I.}~\bibnamefont {Zozoulenko}}, \bibinfo {author}
  {\bibfnamefont {M.}~\bibnamefont {Fahlman}}, \bibinfo {author} {\bibfnamefont
  {P.~J.}\ \bibnamefont {Murphy}}, \bibinfo {author} {\bibfnamefont
  {M.}~\bibnamefont {Berggren}}, \ and\ \bibinfo {author} {\bibfnamefont
  {X.}~\bibnamefont {Crispin}},\ }\href {\doibase 10.1038/nmat3824} {\bibfield
  {journal} {\bibinfo  {journal} {Nature Materials}\ }\textbf {\bibinfo
  {volume} {13}},\ \bibinfo {pages} {1} (\bibinfo {year} {2013})}\BibitemShut
  {NoStop}%
\bibitem [{\citenamefont {Chen}\ \emph {et~al.}(2012)\citenamefont {Chen},
  \citenamefont {Wang},\ and\ \citenamefont {Shuai}}]{Chen2012}%
  \BibitemOpen
  \bibfield  {author} {\bibinfo {author} {\bibfnamefont {J.}~\bibnamefont
  {Chen}}, \bibinfo {author} {\bibfnamefont {D.}~\bibnamefont {Wang}}, \ and\
  \bibinfo {author} {\bibfnamefont {Z.}~\bibnamefont {Shuai}},\ }\href
  {\doibase 10.1021/ct3004436} {\bibfield  {journal} {\bibinfo  {journal}
  {Journal of Chemical Theory and Computation}\ }\textbf {\bibinfo {volume}
  {8}},\ \bibinfo {pages} {3338} (\bibinfo {year} {2012})}\BibitemShut
  {NoStop}%
\bibitem [{\citenamefont {Wang}\ \emph {et~al.}(2012)\citenamefont {Wang},
  \citenamefont {Shi}, \citenamefont {Chen}, \citenamefont {Xi},\ and\
  \citenamefont {Shuai}}]{Wang2012}%
  \BibitemOpen
  \bibfield  {author} {\bibinfo {author} {\bibfnamefont {D.}~\bibnamefont
  {Wang}}, \bibinfo {author} {\bibfnamefont {W.}~\bibnamefont {Shi}}, \bibinfo
  {author} {\bibfnamefont {J.}~\bibnamefont {Chen}}, \bibinfo {author}
  {\bibfnamefont {J.}~\bibnamefont {Xi}}, \ and\ \bibinfo {author}
  {\bibfnamefont {Z.}~\bibnamefont {Shuai}},\ }\href {\doibase
  10.1039/c2cp42710a} {\bibfield  {journal} {\bibinfo  {journal} {Physical
  chemistry chemical physics : PCCP}\ }\textbf {\bibinfo {volume} {14}},\
  \bibinfo {pages} {16505} (\bibinfo {year} {2012})}\BibitemShut {NoStop}%
\bibitem [{\citenamefont {L{\"{u}}ssem}\ \emph {et~al.}(2013)\citenamefont
  {L{\"{u}}ssem}, \citenamefont {Riede},\ and\ \citenamefont
  {Leo}}]{LuessemRiedeLeo2013-PSS}%
  \BibitemOpen
  \bibfield  {author} {\bibinfo {author} {\bibfnamefont {B.}~\bibnamefont
  {L{\"{u}}ssem}}, \bibinfo {author} {\bibfnamefont {M.}~\bibnamefont {Riede}},
  \ and\ \bibinfo {author} {\bibfnamefont {K.}~\bibnamefont {Leo}},\ }\href
  {\doibase 10.1002/pssa.201228310} {\bibfield  {journal} {\bibinfo  {journal}
  {Physica Status Solidi A}\ }\textbf {\bibinfo {volume} {210}},\ \bibinfo
  {pages} {9} (\bibinfo {year} {2013})}\BibitemShut {NoStop}%
\bibitem [{\citenamefont {Fujimori}\ \emph {et~al.}(1994)\citenamefont
  {Fujimori}, \citenamefont {Hoshimono},\ and\ \citenamefont
  {Fujita}}]{Fujimori1994}%
  \BibitemOpen
  \bibfield  {author} {\bibinfo {author} {\bibfnamefont {S.}~\bibnamefont
  {Fujimori}}, \bibinfo {author} {\bibfnamefont {K.}~\bibnamefont {Hoshimono}},
  \ and\ \bibinfo {author} {\bibfnamefont {S.}~\bibnamefont {Fujita}},\ }\href
  {\doibase 10.1016/0038-1098(94)90208-9} {\bibfield  {journal} {\bibinfo
  {journal} {Solid State Communications}\ }\textbf {\bibinfo {volume} {89}},\
  \bibinfo {pages} {437} (\bibinfo {year} {1994})}\BibitemShut {NoStop}%
\bibitem [{\citenamefont {Pfeiffer}\ \emph {et~al.}(1998)\citenamefont
  {Pfeiffer}, \citenamefont {Beyer}, \citenamefont {Fritz},\ and\ \citenamefont
  {Leo}}]{Pfeiffer1998}%
  \BibitemOpen
  \bibfield  {author} {\bibinfo {author} {\bibfnamefont {M.}~\bibnamefont
  {Pfeiffer}}, \bibinfo {author} {\bibfnamefont {A.}~\bibnamefont {Beyer}},
  \bibinfo {author} {\bibfnamefont {T.}~\bibnamefont {Fritz}}, \ and\ \bibinfo
  {author} {\bibfnamefont {K.}~\bibnamefont {Leo}},\ }\href {\doibase
  10.1063/1.122718} {\bibfield  {journal} {\bibinfo  {journal} {Applied Physics
  Letters}\ }\textbf {\bibinfo {volume} {73}},\ \bibinfo {pages} {3202}
  (\bibinfo {year} {1998})}\BibitemShut {NoStop}%
\bibitem [{\citenamefont {Olson}\ \emph {et~al.}(1993)\citenamefont {Olson},
  \citenamefont {Topp},\ and\ \citenamefont {Pohl}}]{Olson1993}%
  \BibitemOpen
  \bibfield  {author} {\bibinfo {author} {\bibfnamefont {J.~R.}\ \bibnamefont
  {Olson}}, \bibinfo {author} {\bibfnamefont {K.~A.}\ \bibnamefont {Topp}}, \
  and\ \bibinfo {author} {\bibfnamefont {R.~O.}\ \bibnamefont {Pohl}},\ }\href
  {\doibase 10.1126/science.259.5098.1145} {\bibfield  {journal} {\bibinfo
  {journal} {Science}\ }\textbf {\bibinfo {volume} {259}},\ \bibinfo {pages}
  {1145} (\bibinfo {year} {1993})}\BibitemShut {NoStop}%
\bibitem [{\citenamefont {Menke}\ \emph
  {et~al.}(2012{\natexlab{a}})\citenamefont {Menke}, \citenamefont {Wei},
  \citenamefont {Ray}, \citenamefont {Kleemann}, \citenamefont {Naab},
  \citenamefont {Bao}, \citenamefont {Leo},\ and\ \citenamefont
  {Riede}}]{Menke2012a}%
  \BibitemOpen
  \bibfield  {author} {\bibinfo {author} {\bibfnamefont {T.}~\bibnamefont
  {Menke}}, \bibinfo {author} {\bibfnamefont {P.}~\bibnamefont {Wei}}, \bibinfo
  {author} {\bibfnamefont {D.}~\bibnamefont {Ray}}, \bibinfo {author}
  {\bibfnamefont {H.}~\bibnamefont {Kleemann}}, \bibinfo {author}
  {\bibfnamefont {B.~D.}\ \bibnamefont {Naab}}, \bibinfo {author}
  {\bibfnamefont {Z.}~\bibnamefont {Bao}}, \bibinfo {author} {\bibfnamefont
  {K.}~\bibnamefont {Leo}}, \ and\ \bibinfo {author} {\bibfnamefont
  {M.}~\bibnamefont {Riede}},\ }\href {\doibase 10.1016/j.orgel.2012.09.024}
  {\bibfield  {journal} {\bibinfo  {journal} {Organic Electronics}\ }\textbf
  {\bibinfo {volume} {13}},\ \bibinfo {pages} {3319} (\bibinfo {year}
  {2012}{\natexlab{a}})}\BibitemShut {NoStop}%
\bibitem [{\citenamefont {Menke}(2013)}]{TorbenMenkeDiss}%
  \BibitemOpen
  \bibfield  {author} {\bibinfo {author} {\bibfnamefont {T.}~\bibnamefont
  {Menke}},\ }\emph {\bibinfo {title} {Molecular Doping of Organic
  Semiconductors -- A Conductivity and Seebeck Study}},\ \href
  {http://www.dr.hut-verlag.de/9783843911771.html} {\bibinfo {type}
  {Dissertation}},\ \bibinfo  {school} {TU Dresden, ISBN~978-3-8439-1177-1,}
  (\bibinfo {year} {2013})\BibitemShut {NoStop}%
\bibitem [{\citenamefont {Menke}\ \emph {et~al.}(2014)\citenamefont {Menke},
  \citenamefont {Ray}, \citenamefont {Kleemann}, \citenamefont {Hein},
  \citenamefont {Leo},\ and\ \citenamefont {Riede}}]{Menke2013}%
  \BibitemOpen
  \bibfield  {author} {\bibinfo {author} {\bibfnamefont {T.}~\bibnamefont
  {Menke}}, \bibinfo {author} {\bibfnamefont {D.}~\bibnamefont {Ray}}, \bibinfo
  {author} {\bibfnamefont {H.}~\bibnamefont {Kleemann}}, \bibinfo {author}
  {\bibfnamefont {M.~P.}\ \bibnamefont {Hein}}, \bibinfo {author}
  {\bibfnamefont {K.}~\bibnamefont {Leo}}, \ and\ \bibinfo {author}
  {\bibfnamefont {M.}~\bibnamefont {Riede}},\ }\href {\doibase
  10.1016/j.orgel.2013.11.033} {\bibfield  {journal} {\bibinfo  {journal}
  {Organic Electronics}\ }\textbf {\bibinfo {volume} {15}},\ \bibinfo {pages}
  {365} (\bibinfo {year} {2014})}\BibitemShut {NoStop}%
\bibitem [{\citenamefont {Menke}\ \emph
  {et~al.}(2012{\natexlab{b}})\citenamefont {Menke}, \citenamefont {Ray},
  \citenamefont {Meiss}, \citenamefont {Leo},\ and\ \citenamefont
  {Riede}}]{Menke2012}%
  \BibitemOpen
  \bibfield  {author} {\bibinfo {author} {\bibfnamefont {T.}~\bibnamefont
  {Menke}}, \bibinfo {author} {\bibfnamefont {D.}~\bibnamefont {Ray}}, \bibinfo
  {author} {\bibfnamefont {J.}~\bibnamefont {Meiss}}, \bibinfo {author}
  {\bibfnamefont {K.}~\bibnamefont {Leo}}, \ and\ \bibinfo {author}
  {\bibfnamefont {M.}~\bibnamefont {Riede}},\ }\href {\doibase
  10.1063/1.3689778} {\bibfield  {journal} {\bibinfo  {journal} {Applied
  Physics Letters}\ }\textbf {\bibinfo {volume} {100}},\ \bibinfo {pages}
  {093304} (\bibinfo {year} {2012}{\natexlab{b}})}\BibitemShut {NoStop}%
\bibitem [{\citenamefont {Wei}\ \emph {et~al.}(2012)\citenamefont {Wei},
  \citenamefont {Menke}, \citenamefont {Naab}, \citenamefont {Leo},
  \citenamefont {Riede},\ and\ \citenamefont {Bao}}]{Wei2012}%
  \BibitemOpen
  \bibfield  {author} {\bibinfo {author} {\bibfnamefont {P.}~\bibnamefont
  {Wei}}, \bibinfo {author} {\bibfnamefont {T.}~\bibnamefont {Menke}}, \bibinfo
  {author} {\bibfnamefont {B.~D.}\ \bibnamefont {Naab}}, \bibinfo {author}
  {\bibfnamefont {K.}~\bibnamefont {Leo}}, \bibinfo {author} {\bibfnamefont
  {M.}~\bibnamefont {Riede}}, \ and\ \bibinfo {author} {\bibfnamefont
  {Z.}~\bibnamefont {Bao}},\ }\href {\doibase 10.1021/ja211382x} {\bibfield
  {journal} {\bibinfo  {journal} {Journal of the American Chemical Society}\
  }\textbf {\bibinfo {volume} {134}},\ \bibinfo {pages} {3999} (\bibinfo {year}
  {2012})}\BibitemShut {NoStop}%
\bibitem [{\citenamefont {Fritzsche}(1971)}]{Fritzsche1971}%
  \BibitemOpen
  \bibfield  {author} {\bibinfo {author} {\bibfnamefont {H.}~\bibnamefont
  {Fritzsche}},\ }\href {\doibase 10.1016/0038-1098(71)90096-2} {\bibfield
  {journal} {\bibinfo  {journal} {Solid State Communications}\ }\textbf
  {\bibinfo {volume} {9}},\ \bibinfo {pages} {1813} (\bibinfo {year}
  {1971})}\BibitemShut {NoStop}%
\bibitem [{\citenamefont {Sherman}\ \emph {et~al.}(1960)\citenamefont
  {Sherman}, \citenamefont {Heikes},\ and\ \citenamefont {Ure}}]{Sherman1960}%
  \BibitemOpen
  \bibfield  {author} {\bibinfo {author} {\bibfnamefont {B.}~\bibnamefont
  {Sherman}}, \bibinfo {author} {\bibfnamefont {R.~R.}\ \bibnamefont {Heikes}},
  \ and\ \bibinfo {author} {\bibfnamefont {R.~W.}\ \bibnamefont {Ure}},\ }\href
  {\doibase 10.1063/1.1735380} {\bibfield  {journal} {\bibinfo  {journal}
  {Journal of Applied Physics}\ }\textbf {\bibinfo {volume} {31}},\ \bibinfo
  {pages} {1} (\bibinfo {year} {1960})}\BibitemShut {NoStop}%
\bibitem [{\citenamefont {Snyder}\ and\ \citenamefont
  {Ursell}(2003)}]{Snyder2003}%
  \BibitemOpen
  \bibfield  {author} {\bibinfo {author} {\bibfnamefont {G.}~\bibnamefont
  {Snyder}}\ and\ \bibinfo {author} {\bibfnamefont {T.}~\bibnamefont
  {Ursell}},\ }\href {\doibase 10.1103/PhysRevLett.91.148301} {\bibfield
  {journal} {\bibinfo  {journal} {Physical Review Letters}\ }\textbf {\bibinfo
  {volume} {91}},\ \bibinfo {pages} {148301} (\bibinfo {year}
  {2003})}\BibitemShut {NoStop}%
\bibitem [{\citenamefont {Tietze}\ \emph {et~al.}(2013)\citenamefont {Tietze},
  \citenamefont {W{\"{o}}lzl}, \citenamefont {Menke}, \citenamefont {Fischer},
  \citenamefont {Riede}, \citenamefont {Leo},\ and\ \citenamefont
  {L{\"{u}}ssem}}]{TietzeMenke2013}%
  \BibitemOpen
  \bibfield  {author} {\bibinfo {author} {\bibfnamefont {M.~L.}\ \bibnamefont
  {Tietze}}, \bibinfo {author} {\bibfnamefont {F.}~\bibnamefont {W{\"{o}}lzl}},
  \bibinfo {author} {\bibfnamefont {T.}~\bibnamefont {Menke}}, \bibinfo
  {author} {\bibfnamefont {A.}~\bibnamefont {Fischer}}, \bibinfo {author}
  {\bibfnamefont {M.}~\bibnamefont {Riede}}, \bibinfo {author} {\bibfnamefont
  {K.}~\bibnamefont {Leo}}, \ and\ \bibinfo {author} {\bibfnamefont
  {B.}~\bibnamefont {L{\"{u}}ssem}},\ }\href {\doibase 10.1002/pssa.201330049}
  {\bibfield  {journal} {\bibinfo  {journal} {Physica Status Solidi A}\
  }\textbf {\bibinfo {volume} {210}},\ \bibinfo {pages} {2188} (\bibinfo {year}
  {2013})}\BibitemShut {NoStop}%
\bibitem [{\citenamefont {Itaka}\ \emph {et~al.}(2006)\citenamefont {Itaka},
  \citenamefont {Yamashiro}, \citenamefont {Yamaguchi}, \citenamefont
  {Haemori}, \citenamefont {Yaginuma}, \citenamefont {Matsumoto}, \citenamefont
  {Kondo},\ and\ \citenamefont {Koinuma}}]{Itaka2006}%
  \BibitemOpen
  \bibfield  {author} {\bibinfo {author} {\bibfnamefont {K.}~\bibnamefont
  {Itaka}}, \bibinfo {author} {\bibfnamefont {M.}~\bibnamefont {Yamashiro}},
  \bibinfo {author} {\bibfnamefont {J.}~\bibnamefont {Yamaguchi}}, \bibinfo
  {author} {\bibfnamefont {M.}~\bibnamefont {Haemori}}, \bibinfo {author}
  {\bibfnamefont {S.}~\bibnamefont {Yaginuma}}, \bibinfo {author}
  {\bibfnamefont {Y.}~\bibnamefont {Matsumoto}}, \bibinfo {author}
  {\bibfnamefont {M.}~\bibnamefont {Kondo}}, \ and\ \bibinfo {author}
  {\bibfnamefont {H.}~\bibnamefont {Koinuma}},\ }\href {\doibase
  10.1002/adma.200502752} {\bibfield  {journal} {\bibinfo  {journal} {Advanced
  Materials}\ }\textbf {\bibinfo {volume} {18}},\ \bibinfo {pages} {1713}
  (\bibinfo {year} {2006})}\BibitemShut {NoStop}%
\bibitem [{\citenamefont {Sumino}\ \emph {et~al.}(2011)\citenamefont {Sumino},
  \citenamefont {Harada}, \citenamefont {Ikeda}, \citenamefont {Tanaka},
  \citenamefont {Miyazaki},\ and\ \citenamefont {Adachi}}]{SuminoHarada2011}%
  \BibitemOpen
  \bibfield  {author} {\bibinfo {author} {\bibfnamefont {M.}~\bibnamefont
  {Sumino}}, \bibinfo {author} {\bibfnamefont {K.}~\bibnamefont {Harada}},
  \bibinfo {author} {\bibfnamefont {M.}~\bibnamefont {Ikeda}}, \bibinfo
  {author} {\bibfnamefont {S.}~\bibnamefont {Tanaka}}, \bibinfo {author}
  {\bibfnamefont {K.}~\bibnamefont {Miyazaki}}, \ and\ \bibinfo {author}
  {\bibfnamefont {C.}~\bibnamefont {Adachi}},\ }\href {\doibase
  10.1063/1.3631633} {\bibfield  {journal} {\bibinfo  {journal} {Applied
  Physics Letters}\ }\textbf {\bibinfo {volume} {99}},\ \bibinfo {pages}
  {093308} (\bibinfo {year} {2011})}\BibitemShut {NoStop}%
\bibitem [{\citenamefont {Harada}\ \emph {et~al.}(2010)\citenamefont {Harada},
  \citenamefont {Sumino}, \citenamefont {Adachi}, \citenamefont {Tanaka},\ and\
  \citenamefont {Miyazaki}}]{Harada2010}%
  \BibitemOpen
  \bibfield  {author} {\bibinfo {author} {\bibfnamefont {K.}~\bibnamefont
  {Harada}}, \bibinfo {author} {\bibfnamefont {M.}~\bibnamefont {Sumino}},
  \bibinfo {author} {\bibfnamefont {C.}~\bibnamefont {Adachi}}, \bibinfo
  {author} {\bibfnamefont {S.}~\bibnamefont {Tanaka}}, \ and\ \bibinfo {author}
  {\bibfnamefont {K.}~\bibnamefont {Miyazaki}},\ }\href {\doibase
  10.1063/1.3456394} {\bibfield  {journal} {\bibinfo  {journal} {Applied
  Physics Letters}\ }\textbf {\bibinfo {volume} {96}},\ \bibinfo {pages}
  {253304} (\bibinfo {year} {2010})}\BibitemShut {NoStop}%
\bibitem [{\citenamefont {Feng-Xing}\ \emph {et~al.}(2008)\citenamefont
  {Feng-Xing}, \citenamefont {Jing-Kun}, \citenamefont {Bao-Yang},
  \citenamefont {Yu}, \citenamefont {Rong-Jin},\ and\ \citenamefont
  {Lai-Feng}}]{FengXing2008}%
  \BibitemOpen
  \bibfield  {author} {\bibinfo {author} {\bibfnamefont {J.}~\bibnamefont
  {Feng-Xing}}, \bibinfo {author} {\bibfnamefont {X.}~\bibnamefont {Jing-Kun}},
  \bibinfo {author} {\bibfnamefont {L.}~\bibnamefont {Bao-Yang}}, \bibinfo
  {author} {\bibfnamefont {X.}~\bibnamefont {Yu}}, \bibinfo {author}
  {\bibfnamefont {H.}~\bibnamefont {Rong-Jin}}, \ and\ \bibinfo {author}
  {\bibfnamefont {L.}~\bibnamefont {Lai-Feng}},\ }\href {\doibase
  10.1088/0256-307X/25/6/076} {\bibfield  {journal} {\bibinfo  {journal}
  {Chinese Physics Letters}\ }\textbf {\bibinfo {volume} {25}},\ \bibinfo
  {pages} {2202} (\bibinfo {year} {2008})}\BibitemShut {NoStop}%
\end{thebibliography}
\end{document}